\documentstyle[prl,aps]{revtex}

\def\beq{\begin{equation}}
\def\eeq{\end{equation}}

\begin{document}
\title{Quantum Superconductor-Metal Transition in a 2D
 Proximity-Coupled Array}
\author{M. V. Feigel'man$^1$ and A. I. Larkin$^{2,1}$}
\address{$^1$ L. D. Landau Institute for Theoretical Physics, \\
Moscow 117940, RUSSIA}
\address{$^2$ Theoretical Physics Institute, University of Minnesota,\\
Minneapolis, MN 55455, USA}
\date{\today}
\maketitle

\begin{abstract}
We construct a theory of quantum fluctuatons in a regular array of small
superconductive islands of size $d$ connected via low-resistance
 tunnel contacts ($G_t = h/4e^2R_t \gg 1$)
to dirty thin metal film with
dimensionless conductance $g\gg 1$. Electrons in the film interact
repulsively with dimensionless strength $\lambda$.
The system is macroscopically superconductive
when distance $b$ between neighbouring islands is short enough.
The zero-temperature phase transition from superconductive to
normal-conductive state is shown to occur with the increase of distance
between superconductive islands, at  $\ln b_c/d \sim G_t^2/\lambda g$.
The critical distance $b_{c}$ is much less than 2d localization length
$L_{loc}\sim e^{\pi g}$,
 so the considered effect develops when
 weak-localization corrections are still small. 
The $T_{c}(g,b)$ dependence at $b<b_{c}$ is found. These results are valid
at sufficiently large $g$, whereas decrease of $g$ is expected to lead
 eventually to another $b_c(g)$ dependence, $\ln b_c/d \sim \sqrt{g}$.

\end{abstract}

\vspace{1cm} {\bf 1. Introduction}. Quantum phase transitions take place at
zero temperature due to variation of the parameters of the system.
 An important
group of examples of such  phase transitions is related to superconductors.
The character of a phase transition here depends on the type of the system.
In  granular superconductors, a phase transition between macroscopically
superconductive and insulating states occurs when the Josephson coupling
energy becomes less than the Coulomb energy due to transfer of a single
Cooper pair between two grains \cite{efetov}. Experiments demonstrating this
type of phase transition in thin superconductive films are numerous \cite
{exp1}. More recently, similar results were obtained on artificially
prepared almost regular Josephson arrays \cite{chalmers,delft1,delft2}.
 A theory of superconductor-insulator phase transition was discussed in \cite
{kivelson,mfisher}. 
A different kind of 2D superconductor - homogeneously
disordered films - was considered by Finkelstein~\cite{fin1}
who predicted superconductive-normal phase transition (vanishing of $T_{c}$) 
due to disorder-enchanced Coulomb repulsion in the Cooper channel
(relevant experimental results are discussed in \cite{fin2}).

Within BCS approximation nonmagnetic impurities do not suppress
the temperature of superconductive transition ("Anderson theorem",  
cf.~\cite{And,AG}). However in the situation when $T_c$ is suppressed
by quantum fluctuations, nonmagnetic impurities can lead to further
suppression of $T_c$. An example of such an effect 
 in quasi-one-dimensional superconductors
was considered by V.I.Mel'nikov and one of the authors~\cite{LM}.
Another example of this kind is studied in the present
paper. We consider  2D superconductive
system where small superconductive (SC) islands are connected via insulating
barriers to a dirty
metal film with good conductance, $g=\sigma \cdot 2\pi\hbar/4e^{2}\gg 1$ 
(below we put $\hbar \equiv 1$). Normal-state tunnel conductance $\sigma_t =
G_t \cdot 4e^2/2\pi\hbar$ of the barrier  is  assumed to be high, 
$G_t \gg 1$.
 We show that in this case
quantum fluctuations destroy superconductive order even at $T=0$, if
distance between neighbouring SC islands $b$ becomes sufficiently long (in
this paper we will assume regular lattice of SC islands). Although the
critical inter-island distance $b_{c}$ appears to be exponentially long at 
$g\gg 1, G_t \gg 1$, it is still much shorter than the 2D localization length $%
L_{loc}\sim \exp (\pi g)$, so the destruction of superconductivity produces
normal-metal state (at least above weak
localization temperature $T_{loc} \sim D/L^2_{loc}$).

Since we consider temperatures much below local $T_{c}^{0}$ of SC islands,
any fluctuations of their gap magnitude $|\Delta |$ can be neglected, and
the ''phase-only'' Hamiltonian is appropriate, with its part responsible for
the inter-island coupling due to proximity effect in the normal conductor
(or due to direct tunnelling as in Josephson junction arrays) given by 
\begin{equation}
H_{int}=-\frac{1}{2}\sum_{ij}J_{ij}\cos (\phi _{i}-\phi _{j})  \label{Hint}
\end{equation}
If the coupling goes via 2D metal, then at $T=0$ and neglecting
electron-electron interaction, one gets (cf. e.g.\cite{ALO}) $J_{ij}\equiv
J(|r_{ij}|)\propto 1/|r_{ij}|^{2}$. Such a very slow decay of coupling
strength allows us to treat the inter-island coupling in the mean-field
approximation (MFA). If the temperature is not sufficiently low,
 then only
nearest-neighbouring islands interact and MFA can be used for an
order-of-magntiude estimates of the transition point. Within MFA the action
of all other islands on a given one is described by the mean ''field'' they
produce: $J\langle \cos \phi \rangle \int dt\cos \phi (t)$. As a result, the
transition point for the appearance of superconductivity at low temperatures
is determined by the condition (cf. e.g. \cite{EL} where a similar transition
in quasi-1D system was considered) 
\begin{equation}
1=\frac{1}{2}J\int_{0}^{1/T}C_{0}(t)dt;\qquad J=\sum_{j}J_{ij}\;;\ \ \
C_{0}(t)=\langle \cos ({\phi (0)-\phi (t))}\rangle =2\langle \cos {\phi (0)}%
\cos {\phi (t)}\rangle 
\label{point}
\end{equation}
where $C_{0}(t)$ is the imaginary-time autocorrelation function describing
phase fluctuations of a single SC island in the metal film. 
These phase fluctuations are determined by the charge transport between
the SC island and surrounding 2D metal.  Since we are going to consider
very low temperatures $T \ll \Delta$ and, moreover, the energy scale of 
all phenomena 
related to establishing the inter-island SC coherence is much below
$\Delta$, the only channel of charge transport between SC island
and 2D metal is the Andreev subgap conduction \cite{andreev}. Corresponding
phase fluctuations can be conviniently described by the imaginary-time
 dissipative
action introduced in \cite{bruder,zaikin}, which is similar to the one 
 of single-electron tunnelling in normal tunnelling junctions
(cf. e.g. \cite{SZ}):
\beq
S_0[\phi(t)] =  - \int dt dt' K^A(t-t') \cos(\phi(t)-\phi(t')) ; \qquad
K^A(t) = \frac{G_A}{(2\pi t)^2}   
\label{SA1}
\eeq
where $G_A$ is the Andreev subgap conductance (normalized to conductance
quantum $4e^2/2\pi\hbar$) between SC island and 2D metal. 
We will always assume that $G_A \gg 1$.

The crucial feature of this dissipative
 action as compared to the standard
dissipative action for Josephson junctions with resistive shunt
\cite{schmid}, is that it is periodic in the phase difference 
$\phi(t)-\phi(t')$, thus it correctly describes the fact that
the electron charge in the island  can be changed by  $2e$ quanta
only. As a result, in the longest-time limit $t \gg t_c$
the correlation function $C_0(t)$ decays sufficiently fast ($\sim 1/t^2$)
and the integral
$\int_{0}^{\infty }C_{0}(t)dt \propto t_c$
 always converges, which is crucial 
(cf. Eq.(\ref{point}) for
the existence of the zero-$T$ quantum phase transition. The "correlation time"
$t_c$ introduced above appears to be exponentially long,
 $t_c \propto e^{G_A/2}$ (cf. \cite{GS,Korsh}). 
It means that the Josephson
coupling must be  exponentially weak in $G_A$ in order to destroy
superconductive coherence in the array at $T=0$.  On the contrary, in the
linear dissipation model \cite{schmid} coherence is preserved at arbitrarily
weak $J$ as long as the dissipative conductance $G_N \geq 1$ (at large $G_N$
the correlation function decays very slowly, $C_0(t) \sim t^{-2/G_N}$ ).

Below we will analize (in Sec.2) the behaviour of Andreev conductance $G_A$
as a function of the normal-state barrier conductance $G_t$ and the sheet
conductance of the metal $g$.  Then we will calculate (in Sec.3)
the low-temperature phase boundary between macroscopically superconductive
and normal-metal states of the array. 
The Sec.4 is devoted to the 
discussion of our results and to the conclusions.

{\bf 2. Andreev conductance and phase correlations of a single island.} 

We start from the case of relatively low transparance of the insulating
barrier between SC island and 2D metal film, i.e., we will assume that
the subgap Andreev conductance $G_A$ can be calculated to the second 
(i.e. lowest  nonzero)
 order in the normal-state tunnelling probability (presise condition
will be discussed below).  Convenient expression for the 
phase-coherent subgap conductance  can be found, e.g., in Eq. (12) of
 paper \cite{nazarov1} :
\beq
G_A(\omega) = \frac{G_t^2}{2\pi g}\ln\frac{\omega_d}{\omega}
\label{GA}
\eeq
where $G_t = 2\pi \hbar/(2e)^2 R_t$ is the dimensionless tunnelling
conductance in the normal state,
 $\omega_d = D/d^2$, with $d$ being size of the SC island, and
 $\omega$ is the frequency of the measurement (for dc conductance,
$\omega$ should be replaced in Eq.(\ref{GA}) by the applied voltage $V$).
Extra (with respect to \cite{nazarov1})
 coefficient 2 in the denominator of Eq.(\ref{GA}) is due to a different
geometry: in our case Cooperon propagates over the whole 2D metallic
 plane, whereas it was half-plane in Eq.(12) of~\cite{nazarov1}.
The physical meaning of the factor $(1/g)\ln\frac{\omega_d}{\Omega}$
in the expression (\ref{GA}) for $G_A$  is that it provides the return
probability for two electrons  to come back to the same point where
they tunnel from the SC island, i.e. it is the time integral over
the Cooperon return amplitude ${\cal C}(r=0,t)$. Below we will always
assume that $G_A(\omega\sim\omega_d) \gg 1$, i.e.
\beq
G_A^0 = \frac{G_t^2}{2\pi g} \gg 1
\label{cond0}
\eeq
Expression (\ref{GA}) was derived assuming no electron-electron
interaction in the normal metal. It can be generalized to the
case of nonzero (dimensionless) repulsion in the Cooper channel 
$\lambda > 0$. To do so, we note following \cite{ALO} that in the
 presence of interaction 
\beq 
\int dt e^{i\omega t} {\cal C}(r=0,t) = \frac{1}{2\pi g}
\frac{\ln\frac{\omega_d}{\omega}}{1+\lambda\ln\frac{\omega_d}{\omega}}
\label{Cooplambda}
\eeq
where 
\begin{equation}
\lambda \equiv \lambda (d)=
\frac{\lambda _{0}}{1+\lambda _{0}\ln\frac{E_{F}}{\omega_d}}
\label{lambda}
\end{equation}
is an effective dimensionless constant of electron-electron repulsion in the
metal (defined on the scale $d$).
Therefore at sufficiently low frequencies 
$\omega \leq \omega_{\lambda} = \omega_d e^{-1/\lambda}$ the subgap conductance
becomes frequency-independent,
\beq
G_A = \frac{G_t^2}{2\pi g\lambda}, \qquad \omega \ll \omega_{\lambda}
\label{GA2}
\eeq
The results (\ref{GA},\ref{GA2}) for subgap conductivity 
were derived under the assumption that the superconductive phase $\phi$ of
 the island does not fluctuate ( due to the coupling to external
 low-impedance circuits).  In our problem we deal with an array of many
small SC islands which are coupled to the surrounding only by the Andreev
conduction processes, therefore we should take phase fluctuations 
into account explicitely.
We will calculate, using the dissipative action of the type of (\ref{SA1}),
 the single-island correlation function
$C_0(t)$ as defined in Eq.(\ref{point}) and the effective charging 
energy $E^*_C$ defined via the
relation
\beq
1/E^*_C =
 \int_{0}^{\infty}C_{0}(t)dt 
\label{EC}
\eeq
We start from the assumption that the interaction constant $\lambda$ is 
not very small,
so the relevant frequency scale $\omega \ll \omega_{\lambda} $, and 
Andreev conductance $G_A$ is frequency-independent and given by 
Eq.(\ref{GA2})
The corresponding action is given by Eq.(\ref{SA1}), which is formally
equivalent to the problem considered by Kosterlitz~\cite{Kosterlitz}
and then later on by others~\cite{GS,SZ,ZP,S3}. 
The bare correlation function for the phase fluctuations as determined by 
Eq.(\ref{SA1}) is given by
\beq
\langle|\phi_{\omega}|^2\rangle =  \frac{2\pi}{|\omega|G_A}
\label{phicorr}
\eeq
and leads to the logarithmic growth of phase fluctuations in the
time domain at $t \gg\omega_{\lambda}^{-1}$ (upper limit of this region
will be discussed below),
\beq 
S_0(t) = \frac{1}{2}\langle (\phi(t)-\phi(0))^2\rangle_0 = 
(2/G_A)\ln\omega_{\lambda}t 
\label{S0}
\eeq
Now one could calculate $C_0(t)$ in Gaussian approximation using
 Eq.(\ref{S0}) and get power-law decay with the exponent $2/G_A$; however,
such a procedure would be incorrect since the parameter $G_A$ gets 
renormalized due to the nonlinearity of the action (\ref{SA1}).  The 
renormalization group (RG) equations to the main order in $1/G_A$
 were derived in~\cite{Kosterlitz}:
\beq
\frac{dG_A(\zeta)}{d \zeta} = - 2; \qquad 
\frac{d\ln C_0(\zeta)}{d \zeta} = - \frac{2}{G_A(\zeta)}
\label{RG}
\eeq
where $\zeta=\ln(\omega_{\lambda}/\omega)$. 
These equations can be trivially integrated, the solution is
\beq
\tilde{G}_A(\omega) = G_A - 2\ln\frac{\omega_{\lambda}}{\omega}; \quad 
C_0(t) = 1 - \frac{2}{G_A}\ln\omega_{\lambda}t
\label{RGSol}
\eeq
Physical meaning of $\tilde{G}_A(\omega)$ dependence as given by 
Eq.(\ref{RGSol}) can be understood as follows: due to the fluctuations of 
$\phi(t)$, the coherence between successive Andreev reflections of the
electron which have been diffusing in the metal during long time 
$\sim 1/\omega$, is partially destroyed, which leads to the decrease of
subgap conductance on the corresponding timescale.
Note that the Eqs.(\ref{RG},\ref{RGSol}) are valid
as long as renormalized Andreev conductance is large, $G_A(\zeta) \gg 1$,
i.e. $C_0(t)$ decreases according to Eq.(\ref{RGSol}) till 
$t \approx t_c = \omega_{\lambda}^{-1} e^{G_A/2}$, where $C_0(t)$ becomes
of order $1/G_A \ll 1$.  At longer timescales $t \gg t_c$ the correlation
function $C_0(t) \sim G_A^{-1}(t_c/t)^2$. The main contribution to the
integral 
$\int_{0}^{\infty}C_{0}(t)dt = 1/E_C^*$ comes from $t\sim t_c$. Using
(\ref{RGSol}) we obtain the estimate
$E^*_C \approx\frac{G_A}{t_c} \approx \omega_d e^{-1/\lambda} G_Ae^{-G_A/2}$.
The above evaluation of $t_c$ was based on the leading order of RG approach,
thus it does not, in general, give  the correct preexponential factor.
A more accurate calculation of $E^*_C$ in a similar problem was done in
\cite{ZP} with the use of the instanton technique developed in \cite{Korsh}.
The result of~\cite{ZP} (confirmed later on in the Monte Carlo simulation~\cite{S3})
can be translated in our case as
\beq
   E^*_C = {\cal A} \omega_d e^{-1/\lambda}
\left(\frac{G_A}{\pi}\right)^2e^{-G_A/2} =
    {\cal A} \omega_d e^{-1/\lambda}
\left(\frac{G_t^2}{2\pi^2g\lambda}\right)^2e^{-G_t^2/4\pi g\lambda} 
\label{EC1}
\eeq
where ${\cal A}$ is some numerical factor of order 1. 
Thus the effective Coulomb energy $E_C^*$
 associated with the Andreev conduction process  is still nonzero, although
 exponentially
 suppressed at large $G_A$ (basically the same result was mentioned in 
\cite{zaikin}).  This is why we expect quantum ( at $T=0$) phase transition 
to occur when the Josephson coupling energy $J$ is of the order of $E_C^*$.

Before we turn to the discussion of the phase transition, let us
analize phase fluctuations in the frequency regions where {\it bare} Andreev
conductance differs from (\ref{GA2}), and is frequency-dependent 
(cf. Eqs.(\ref{GA}).
The corresponding dissipative action is of the form (\ref{SA1}), but with
a generalized expression for the kernel $K^A(t)$, which is defined via
\beq
 \int dtK^A(t) (e^{i\omega t}-1) = - \frac{|\omega|}{4\pi}G_A(\omega) 
\label{KAnew}
\eeq
Now,  Eq.(\ref{phicorr}) will be the same but with frequency-dependent
$G_A(\omega)$, which leads to the following results for $S_0(t)$:
\beq
S_0(t) =  \frac{4\pi g}{G_t^2}\ln\ln\omega_d t; 
\qquad \omega_d^{-1} \ll t \ll \omega_{\lambda}^{-1} \\ 
\label{S2}
\eeq
In the shortest time window $t \leq \omega_{\lambda}^{-1}$ fluctuations
grow extremely slowly, and can be neglected
altogether due to the smallness of the prefactor (cf. Eq.(\ref{cond0})).

{\bf 3. Coupling due to proximity effect and the SC-M transition}. 

Total action of the system contains both single-island dissipative terms 
determining phase dynamics of individual islands,
and Josephson coupling between phases on different islands              
\beq
S = - \int dt dt'\sum_j\left[ K^A(t-t') \cos(\phi_j(t)-\phi_j(t')) -
\int dt dt' \sum_{j\neq k} J_{jk}(t-t') \cos(\phi_j(t)-\phi_k(t'))
\right]
\label{action}
\eeq
As was explained in the Introduction, we are going to look for the
SC-M phase transition within the mean field approximation, which is possible
due to slow decay of 
the Josephson coupling energy $\int J_{jk}(t) dt = J(r_{jk})$ with 
island separation $r_{jk}$.
Josephson energy can be found
(in the leading order in $G_t$)
with the use of the results \cite{ALO} (note that in that paper planar
geometry of S-N-S contact was considered, whereas now we deal with the
cylindrical geometry). Then we come to the following expression for the
Fourrier component of the coupling ${\cal J}(q)$: 
\begin{equation}
{\cal J}(q)= \frac{1}{2\pi^2}\frac{G_{t}^{2}}{\nu }
\frac{\Pi (q)}{1+\lambda \Pi (q)};\quad
\Pi (q)= 
2\pi T\sum_{\omega _{n}}\frac{1}{2|\omega _{n}| + Dq^{2}}=
\ln \frac{\omega _{d}}{max(T,Dq^{2})}
  \label{J}
\end{equation}
 Note that neglecting electron-electron
interaction $\lambda $ one would get logarithmic divergence of ${\cal J}(0)$
in the zero-$T$ limit. The expression (\ref{J}) for the Josephson coupling
 is analogous to Eq.(\ref{GA2}) for the subgap conductivity in the presence
of interaction $\lambda$.
 We will always assume $\lambda \ln \frac{\omega _{d}}{T}\gg 1 $.
 In the real space, Josephson interaction is given (at $T=0$) by
\begin{equation}
J(r) = \frac{G_{t}^{2}}{2\pi^3\nu}\frac{1}{r^2(1+ 2\lambda \ln\frac{r}{d})^2} 
 \label{J0}
\end{equation}
We see that $J(r)$ does not depend on the metal conductance $g$. We need to
find the sum $J = \sum_{{\bf r}_n}J(r_n)$ over the lattice of islands.
This sum can be approximated by the integral over $d^2r$ with the lower
limit $r_{min} \sim b$:
\beq
J = \int_b^\infty\frac{2\pi r dr}{b^2}J(r) = 
\frac{G_t^2}{4\pi^2\nu\lambda^2}\frac{1}{b^2\ln\frac{b}{d}}
\label{J1}
\eeq 
Using now Eqs.(\ref{point},\ref{EC},\ref{EC1},\ref{J1}), 
we obtain the  critical value
of inter-island distance $b_{c}$ such that at $b > b_c$ macroscopically
 superconductive state is absent even at $T=0$:
\begin{equation}
\frac{b_c}{d} = \frac{\pi}{\sqrt{2{\cal A}}}\frac{2\pi g\sqrt{\lambda}}{G_t^2}
\exp\left[ \frac{1}{2\lambda}\left(\frac{G_t^2}{4\pi g} + 1\right)\right]
 \label{b}
\end{equation}
The above formula for $b_c$  may look a bit paradoxical, since it shows
that the critical distance between SC islands {\it grows} with the
decrease of the film conductance $g$. Of course, such a result
is valid in the intermediate range of not very small $g$,
where both $E_C^*$ and 
$J$ can be calculated in the leading order in $G_t$.
Now we discuss the conditions for such a procedure to be valid.

 An important condition for the validity  of Eqs.(\ref{RGSol},\ref{EC1})
comes from consideration of the low-frequency region.
First of all,
the expressions (\ref{GA},\ref{GA2}) are written in the lowest order
in $G_t$ and thus  the necessary condition for their validity consists in
the smallness of the anomalous quasiclassical Green function,
$F({\bf r},\omega) \ll 1$
 everythere in the normal metal, 
including the vicinity of the tunnel barrier $ r = d$. If this condition
(which is equivalent to the inequality $G_A \ll G_t$) is not fulfilled,
 the simple perturbative theory in terms of tunneling amplitude becomes
invalid, and a full nonlinear treatment based
on the Usadel equations for the Green functions 
$F({\bf r},\omega)$ and $ G({\bf r},\omega)$
 must be developed (see e.g.~\cite{nazarov2} and
references therein). 
Thus the  necessary condition can be written as
\beq
\lambda \gg \frac{G_t}{2\pi g} 
\label{cond1}
\eeq
which must be fulfilled together with Eq.(\ref{cond0}).

There is also another assumption implicitely used in the
 derivation of Eq.(\ref{b}). Namely, we neglected
Coulomb-induced contribution \cite{fin1} to the Cooperon-channel 
coupling constant $\lambda$, which is of the order of
$\delta\lambda \sim (1/g)\ln\frac{\omega_d}{\omega}$.
The estimate for $\delta\lambda$ may be derived also along the lines
of Ref. 27, where Coulomb correction $\delta\Pi$ to the Cooperon amplitude
was calculated in the main order over $1/g$. The results of \cite{RZ},
considered in the limit $T\to 0$, shows that the relative correction
\beq
\frac{\delta\Pi (\omega)}{\Pi(\omega)} \sim 
\frac{1}{g}\ln^2\frac{\omega_d}{\omega}.
\label{RZ}
\eeq
Comparison between Eq.(\ref{RZ}) and Eq.(\ref{Cooplambda}) shows
again that Coulomb corrections can be neglected as long as
\beq
\lambda \gg \frac{1}{g}\ln\frac{\omega_d}{\omega}
\label{est2}
\eeq
  Relevant scale
of frequency is given by $t_c^{-1} = \omega_{\lambda}e^{-G_A/2}$,
so the estimate  Eq.(\ref{est2}) actually coinsides with the condition
given by Eq.(\ref{cond1}).
Therefore the inequility (\ref{cond1}) is needed
for both our assumptions (weakness of tunnelling and of Coulomb
interaction in the metal) to be valid.

Currently we can only speculate about the opposite case
$\lambda \leq G_t/2\pi g$, which is necessarily realized as $g$
decreases. It is clear from the previous estimates, that in this case
both Coulomb interaction in the metal and higher orders in tunnelling
(usually handled by means of the Usadel equation) 
should be taken into acocunt.
 For the purely quialitative discussion, let us first put
$\lambda =0$ and consider the results \cite{nazarov2}
for the subgap conductance beyond the
perturbation expansion in $G_t$. According to \cite{nazarov2}, the
behaviour of the total 
subgap conductance $G_A(\omega)$ is determined by the product 
$G_t r_{2D}$, where $r_{2D}$
 is the total resistance
of the diffusive part of the conducting circuit. 
In our case it is the resistance  of the metal film measured between the
 SC island and the second contact
assumed to be big circle of radius $R \gg d$.
Thus $r_{2D} = (1/2\pi g)\ln\frac{L(\omega,R)}{d}$,
there $L(\omega,R) = \min(\sqrt{D/\omega},R)$.
If $G_t r_{2D} \ll 1$, we are back to the result given by Eq.(\ref{GA}).
In the opposite limit $G_t r_{2D} \gg 1$ the total
subgap conductance was found to be determined by the conductance of the
 metal film only
and is insensitive to the barrier transparency:
\beq
G_A(\omega) = \frac{2\pi g}{\ln\frac{L(\omega,R)}{d}}, 
\label{GA3}
\eeq
Note that in the regime (\ref{GA3}) the subgap conductance decreases with
the decrease of sheet conductance $g$. Comparing it with
Eq.(\ref{GA2}) we conclude that $G_A(g)$ passes via maximum as $g$ decreases.
Using Eq.(\ref{phicorr}) together with Eq.(\ref{GA3}), we can find, at $\lambda =0$ 
and sufficiently low frequencies, that phase fluctuations grow as
\beq
S_0(t) = \frac{1}{4\pi g}\ln^2\omega_dt
\label{S3}
\eeq
(note the similarity of Eq.(\ref{S3})
with the expression for the tunnelling action \cite{LS} describing
zero-biased anomaly of tunnelling conductance into a dirty 2D metal 
\cite{AA}).
Now let us come back to the discussion of $b_c(g)$ dependence.
 We expect that at low enough $g$  critical distance $b_c$
will scale as $\ln b_c/d \propto \sqrt{g}$. This estimate~\cite{fei}
 follows from
Eq.(\ref{S3}) and is analogous to  Finkelstein's result~\cite{fin1}
$g_c \sim 1/\lambda_{BCS}^2 = \ln^2(\omega_D/T_c^0)$
 for vanishing of $T_c$ in a homogenously disordered film. 
Let us now "turn on" repulsion constant $\lambda > 0$, but assume that
it is small, so the inequility opposite to Eq.(\ref{cond1}) is obeyed.
As long as $\lambda$ is also less than the relevant scale $g^{-1/2}$ of the
inverse logarithm in (\ref{GA3}), the result valid
for $\lambda=0$ holds.  However, there exists additional, intermediate
case when $g^{-1/2} \leq \lambda \leq G_t/2\pi g$. Here we expect
the logarithm in (\ref{GA3}) to be replaced by $\lambda^{-1}$,  and therefore
 $G_A \sim g\lambda$.  Thus we arrive at the following picture of the
 ${\cal R}(g) \equiv \ln b_c(g)/d$ evolution: it grows with $g$ decrease 
according 
to Eq.(\ref{b}), untill $g$ reaches $g_1 \approx G_t/2\pi\lambda$ where
${\cal R}(g)$ passes via the maximum value of order $G_t/4$, and then
decreases as ${\cal R}(g) \sim g\lambda$ in the region 
$\lambda^{-2} \leq g \leq G_t/2\pi\lambda $.
 Finally, at lowest $g\leq  \lambda^{-2}$ it decreases as
 ${\cal R}(g) \sim\sqrt{g}$.

In all cases considered, ${\cal R}(g) \ll g$,  which means that
 weak localization
corrections to the conductance of 2D metal~\cite{GLK,AA2} are indeed
 weak even at  $b \geq b_c$. Therefore,
upon destruction of superconductivity  with increase of inter-island 
separation, the normal metal ground-state will be produced.

Now we turn to the discussion of the low-temperature phase boundary
between  superconductive and metal states (assuming that the condition
(\ref{cond1}) is fulfilled, and, also, the temperature $T$ is
so low that single-electron transport via NS barriers can be neglected).
We calculate the dependence $T_c(b)$ of the critical temperature on the
 separation between islands, assuming it is much less than the zero-$T$ 
critical value $b_c$ from Eq.(\ref{b}). 
As a result, we obtain  values of $T_c$ which are higher than
the effective charging energy $E_C^*$ defined in Eq.(\ref{EC1}).
 On the other hand, we will always assume $T_c \ll \omega_{\lambda}$.
Then the MFA transition temperature is determined by the equation
\beq
 J^{-1}_T = \frac{1}{2}\int_0^{1/T} C_0(t) dt = 
\frac{1}{T}\left( 1- \frac{2}{G_A}\ln\frac{\omega_{\lambda}}{T}\right) 
\equiv \frac{1}{T^{eff}}
\label{Tc1}
\eeq
The strength of Josephson coupling depends on temperature (cf.Eq.(\ref{J})),
and this dependence must be taken into account when calculating the
 sum over islands $J_T$, which replaced $J$ from Eq.(\ref{J1}) at $T > 0$.
At lower enough temperatures
 this effect can be accounted for by simply
 choosing the upper limit of integration in Eq.(\ref{J1}) to be the
thermal coherence length $\xi_T = \sqrt{D/T}$. This results in
\beq
J_T = J \frac{\ln\frac{\xi_T}{b}}{\ln\frac{\xi_T}{d}}; \qquad b \ll \xi_T
\label{JT1}
\eeq
Solving Eq.(\ref{Tc1}) together with Eq.(\ref{JT1}) we find 
\beq
T^{MFA}_c(b) = \frac{D}{b^2}\frac{\ln\frac{b_c}{b}}{\lambda\ln^2\frac{b}{d}}
\label{Tmfa}
\eeq 
 This formula can be applied as long as $\xi_T \gg b$, i.e., 
when $T_c^{MFA}(b) \ll D/b^2$, which is
equivalent to the condition $b \gg b_1$ with
\beq
\ln\frac{b_1}{d} \approx \frac{G_t}{\lambda\sqrt{8\pi g}}
\label{b1}
\eeq

At shorter $b$, transition temperature grows in such a way that $\xi_T$
becomes shorter than $b$, and the
 MFA equation (\ref{Tc1})  cannot be used anymore.
In this case an appropriate approximation will be to neglect all couplings 
between non-nearest neighbours, i.e.
 to consider  Berezinsky-Kosterlitz-Thouless (BKT)
phase transition in 2D XY model with interaction between nearest 
neighbours~\cite{Berez}. Moreover, in this region quantum phase fluctuations
becomes weak and can be neglected in determination of $T_c$.
Therefore we estimate transition temperature as
$T_c \approx J(b,T_c)$,
 where the nearest-neighbours Josephson coupling
 $J(b,T)$ in the relevant temperature range can  be evaluated as
\beq
J(b,T) \approx \frac{G_t^2}{8\pi^3\nu}\frac{1}{\xi_T^2}
\frac{\exp(-b/\xi_T)}{\lambda^2\ln^2\frac{\xi_T}{d}} ; \qquad b \gg \xi_T
\label{JT2}
\eeq
 Using Eq.(\ref{JT2})
 and solving the equation $T_c = J(b,T_c)$
for $T_c$ with logarithmic accuracy, we finally obtain
\beq
T_c(b) \approx \frac{D}{b^2}
\ln^2\left(\frac{G_t^2}{16\pi^2 g\lambda^2\ln^2\frac{b}{d}} \right) \qquad
d e^{1/\lambda} \ll b \ll b_1.
 \label{Tc}
\eeq
The expressions (\ref{Tmfa}) and (\ref{Tc}) approximately
 match each other at $b \approx b_1$ defined by Eq.(\ref{b1}).
The phase transition is of mean-field kind at $b \gg b_1$ whereas
it is of the BKT type~\cite{Berez} at $b \ll b_1$.
Therefore the temperature dependencies of macroscopic superconductive 
 parameters  such as
 superfluid density $\rho_s(T)$ and critical current density
 $j_c(T)$ are expected to differ significantly depending on whether $b$
is small or large with respect to $b_1$.

{\bf 4. Conclusions.} We have shown that in a mixed superconductor-normal
system with a small fraction of superconductive regions quantum fluctuations
effects still persist, and lead to the destruction of superconductivity even
at $T=0$, in spite of the good metal sheet conductance $\sigma \gg 4e^2/h$
and the low  SC-M interface resitance $R_t \ll h/4e^2$.
 These effects decay exponentially with the Andreev subgap
conductance $G_A = G_t^2/2\pi g\lambda \gg 1$;
 in particular, macroscopic superconductivity
persists down to the  relative fraction 
of SC material $n_{sc} \approx (d/b_c)^2 \sim \exp(-G_A/2-1/\lambda)$ .
The temperature scale relevant for the effect we discussed is about 
$T_1 = D/b_c^2$.  At the distances between neighboring islands $b \ll b_c$
the SC-M transition line  $T_c(b,g)$ is determined by 
 Eq.(\ref{Tmfa}) and Eq.(\ref{Tc}).
These results were derived under  the
assumption that the Andreev subgap resistance is mainly determined by the S-N
boundary. With the decrease of  $g$, the subgap resistance  first decreases
 but then starts to grow when the resistance of
 the film dominates the S-N boundary resistance.
Quantitative investigation of quantum fluctuations and phase diagram
 in the corresponding region of parameters 
(the condition opposite to Eq.(\ref{cond1}))
will be a subject of future work;  we expect that at smallest 
$g \leq \lambda^{-2}$ the universal result
$-\ln n_{sc} \propto \sqrt{g}$ holds.
 It is important that the considered effects of
quantum phase fluctuations are strong in the
temperature range there usual interference and interaction-induced
corrections to conductivity are still weak. 
In the normal metal state and at relatively high temperatures these
fluctuations are expected to produce logarithmic in temperature corrections
to conductivity, with the magnitude considerably larger than the
weak-localization ones.

There are several open problems closely related to the simple model we have
studied here.  It would be interesting to study the influence of
quantum (T=0) fluctuations of the phases $\phi_j(t)$ in the metallic phase
 ( $b > b_c$) on weak localization corrections to metal conductance and to
check if they produce finite "dephasing time" $\tau_{\phi}(0)$.
Strong dependence of the effective Coulomb energy $E_C^*$ on the film
conductance indicate an importance of mesoscopic fluctuations of conductance
between different regions of nominally uniform film. 
Intersting problem is related to the effect of magnetic field of the scale 
$B \sim \Phi_0/b^2$, which
may lead to a formation of a superconductive glass state
(cf. e.g. \cite{87}). Another important problem is the
 magnetic field - induced SC-normal transition in a uniformly disordered 
SC film at low $T$: it was
argued by Spivak and Zhou \cite{spivak}, that mesoscopic fluctuations leads
to a finite probability to find small SC cluster under the field which
is much above "mean" $H_{c2}$. The problem of phase coherence between such
clusters would resemble our model under the external magnetic field, but
with important modifications due to randomness in the cluster
positions.

We are grateful to I. L. Aleiner, B. L. Altshuler, 
M. E. Gershenzon, V. B. Geshkenbein,
 D. Esteve, L. B. Ioffe, L. I. Glazman, A. S. Ioselevich, G. B. Lesovik, 
Yu. V. Nazarov, H. Pothier, B. Z. Spivak and Fei Zhou for many instructive
discussions. The research of M.V.F. was supported by INTAS-RFBR grant \#
95-0302, Swiss National Science Foundation collaboration grant \# 7SUP
J048531, DGA grant \# 94-1189, and the Programm "Statistical Physics"
from the Russian Ministry of Science.


\begin{references}
\bibitem{fei}  Similar result was obtained recently by Fei Zhou (private
communication).
\end{references}
\end{document}